\documentclass[mathpazo]{cicp}

\usepackage{graphicx}
\usepackage{amssymb}
\usepackage{amsmath}

\begin{document}
 
\title{SPEEDUP Code for Calculation of Transition Amplitudes via the Effective Action Approach}

\author[Bala\v z A et.~al.]{Antun Bala\v z\corrauth, 
Ivana Vidanovi\' c, Danica Stojiljkovi\' c, Du\v san Vudragovi\' c, Aleksandar Beli\' c, Aleksandar Bogojevi\' c}
\address{Scientific Computing Laboratory, Institute of Physics Belgrade,\\
University of Belgrade, Pregrevica 118, 11080 Belgrade, Serbia\\
http://www.scl.rs/}
 \email{{\tt antun@ipb.ac.rs} (A.~Bala\v z)}

\begin{abstract}
We present Path Integral Monte Carlo C code for calculation of quantum mechanical transition amplitudes for 1D models. The SPEEDUP C code is based on the use of higher-order short-time effective actions and implemented to the maximal order $p$=18 in the time of propagation (Monte Carlo time step), which substantially improves the convergence of discretized amplitudes to their exact continuum values. Symbolic derivation of higher-order effective actions is implemented in SPEEDUP Mathematica codes, using the recursive Schr\" odinger equation approach. In addition to the general 1D quantum theory, developed Mathematica codes are capable of calculating effective actions for specific models, for general 2D and 3D potentials, as well as for a general many-body theory in arbitrary number of spatial dimensions.
\end{abstract}

\pac{05.30.-d, 02.60.-x, 05.70.Fh, 03.65.Db}

\maketitle

\section{Introduction}
\label{sec:intro}

Exact solution of a given many-body model in quantum mechanics is usually expressed in terms of eigenvalues and eigenfunction of its Hamiltonian
\begin{equation}
\hat H=\sum_{i=1}^M \frac{\hat{\mathbf p}^2_i}{2m_i}+\hat V(\hat{\mathbf q}_1,\ldots, \hat{\mathbf q}_M)\, ,
\end{equation}
but it can be also expressed through analytic solution for general transition amplitude $A(\mathbf a, \mathbf b;T)=\langle \mathbf b|e^{-iT\hat H/\hbar}|\mathbf a\rangle$ from the initial state $|\mathbf a\rangle$ to the final state $|\mathbf b\rangle$ during the time of propagation $T$. Calculation of transition amplitudes is more suitable if one uses path integral formalism \cite{feynman, feynmanhibbs, kleinertbook}, but in principle, if eigenproblem of the Hamiltonian can be solved, one should be able to calculate general transition amplitudes, and vice versa. However, mathematical difficulties may prevent this, and even more importantly, exact solutions can be found only in a very limited number of cases. Therefore, use of various analytic approximation techniques or numerical treatment is necessary for detailed understanding of the behavior of almost all models of interest.

In numerical approaches it could be demanding and involved to translate numerical knowledge of transition amplitudes to (or from) eigenstates, but practically can be always achieved. It has been implemented in various setups, e.g. through extraction of the energy spectra from the partition function \cite{feynmanhibbs, kleinertbook, danicapla, ivanapla}, and using the diagonalization of space-discretized matrix of the evolution operator, i.e. matrix of transition amplitudes \cite{sethia1990, sethia1999, ivanapre1, ivanapre2, becpla}. All these applications use the imaginary-time formalism  \cite{feynmanstat, parisi}, typical for numerical simulations of such systems.

Recently introduced effective action approach \cite{prl-speedup, prb-speedup, pla-euler, pre-ideal, pre-recursive} provides an ideal framework for exact numerical calculation of quantum mechanical amplitudes. It gives systematic short-time expansion of amplitudes for a general potential, thus allowing accurate calculation of short-time properties of quantum systems directly, as has been demonstrated in Refs.~\cite{ivanapre1, ivanapre2, becpla}. For numerical calculations that require long times of propagation to be considered using e.g. Monte Carlo method, effective action approach provides improved discretized actions leading to the speedup in the convergence of numerically calculated discretized quantities to their exact continuum values. This has been also demonstrated in Monte Carlo calculations of energy expectation values using the improved energy estimators \cite{jelapla, ivanapla}.

In this paper we present SPEEDUP codes \cite{speedup} which implement the effective action approach, and which were used for numerical simulations in Refs.~\cite{danicapla, ivanapla, ivanapre1, ivanapre2, becpla, prl-speedup, prb-speedup, pla-euler, pre-ideal, pre-recursive}. The paper is organized as follows. In Section \ref{sec:theory} we briefly review the recursive approach for analytic derivation of higher-order effective actions. SPEEDUP Mathematica codes capable of symbolic derivation of effective actions for a general one- and many-body theory as well as for specific models is described in detail in Section \ref{sec:Mathematica}, while in Section \ref{sec:C} we describe SPEEDUP Path Integral Monte Carlo C code, developed for numerical calculation of transition amplitudes for 1D models. Section \ref{sec:conclusions} summarizes presented results and gives outlook for further development of the code.

\section{Theoretical background}
\label{sec:theory}

From inception of the path integral formalism, expansion of short-time amplitudes in the time of propagation was used for the definition of path integrals through the time-discretization procedure \cite{feynmanhibbs, kleinertbook}. This is also straightforwardly implemented in the Path Integral Monte Carlo approaches \cite{ceperley}, where one usually relies on the naive discretization of the action. Several improved discretized actions, mainly based on the Trotter formula and its generalizations, were developed and used in the past \cite{takahashiimada, libroughton, deraedt2}. A recent analysis of this method can be found in Jang et al \cite{jangetal}. Several related investigations dealing with the speed of convergence have focused on improvements in short-time propagation \cite{makrimiller,makri} or the action \cite{alfordetal}. More recently, split-operator method has also been developed \cite{chinkrotscheck, hernandez, ciftja, sakkos, janecek}, later extended to include higher-order terms  \cite{bandrauk, chinchen, omelyan, bayepre}, and systematically improved using the multi-product expansion \cite{chinarxiv, krotscheck, chin}.

The effective action approach is based on the ideal discretization concept \cite{pre-ideal}. It was introduced first for single-particle 1D models \cite{prl-speedup, prb-speedup, pla-euler} and later extended to general many-body systems in arbitrary number of spatial dimensions \cite{ivanapla, pre-recursive}. This approach allows systematic derivation of higher-order terms to a chosen order $p$ in the short time of propagation.

Recursive method for deriving discretized effective actions, established in Ref.~\cite{pre-recursive}, is based on solving the underlying Schr\" odinger equation for the amplitude. It has proven to be the most efficient tool for treatment of higher-order expansion. In this section we give brief overview of the recursive method, which will be implemented in Mathematica in the next section. We start with the case of single particle in 1D, used in the SPEEDUP C code. Throughout the paper we will use natural system of units, in which $\hbar$ and all masses are set to unity.

\subsection{One particle in one dimension}
\label{sec:P1D1}

In the effective action approach, transition amplitudes are expressed in terms of the ideal discretized action $S^*$ in the form
\begin{equation}
A(a, b; T)=\frac{1}{\sqrt{2\pi T}}\, e^{-S^*(a, b; T)}\, ,
\end{equation}
which can be also seen as a definition of the ideal action \cite{pre-ideal}. Therefore, by definition, the above expression is correct not only for short times of propagation, but for arbitrary large times $T$. We also introduce the ideal effective potential $W$,
\begin{equation}
S^*(a, b; T)=T\left[\frac{1}{2}\left(\frac{b-a}{T}\right)^2+W\right]\, ,
\end{equation}
reminiscent of the naive discretized action, with the arguments of the effective potential ($a$, $b$, $T$) usually written as $W\left(\frac{a+b}{2}, \frac{b-a}{2}; T\right)$, to emphasize that we will be using mid-point prescription.

However, ideal effective action and effective potential can be calculated analytically only for exactly solvable models, while in all other cases we have to use some approximative method. We use expansion in the time of propagation, assuming that the time $T$ is small. If this is not the case, we can always divide the propagation into $N$ time steps, so that $\varepsilon=T/N$ is small. Long-time amplitude is than obtained by integrating over all short-time ones,
\begin{equation}
A(a, b; T)=\int dq_1\cdots dq_{N-1}\ A(a, q_1; \varepsilon)\,
A(q_1, q_2; \varepsilon)\cdots A(q_{N-1}, b; \varepsilon)\, ,
\label{eq:AMC}
\end{equation}
paving the way towards Path Integral Monte Carlo calculation, which is actually implemented in the SPEEDUP C code.

If we consider general amplitude $A(q,q';\varepsilon)$, introduce the mid-point coordinate $x=(q+q')/2$ and deviation $\bar x=(q'-q)/2$, and express $A$ using the effective potential,
\begin{equation}
A(q, q'; \varepsilon)=\frac{1}{\sqrt{2\pi \varepsilon}}\, e^{-\frac{2}{\varepsilon}\bar x^2-\varepsilon W(x, \bar x; \varepsilon)}\, ,
\end{equation}
the time-dependent Schr\" odinger equation for the amplitude leads to the following equation for $W$
\begin{equation}
W+\bar x\,\bar\partial W+\varepsilon\,\partial_\varepsilon W
-\frac{1}{8}\,\varepsilon\,\partial^2 W-
\frac{1}{8}\,\varepsilon\,\bar\partial^2 W
+\,\frac{1}{8}\,\varepsilon^2\,(\partial W)^2
+\frac{1}{8}\,\varepsilon^2\,(\bar\partial W)^2= \frac{1}{2}\, (V_++V_-)\, ,
\label{eq:eqW}
\end{equation}
where $V_\pm=V(x\pm\bar x)$, i.e. $V_-=V(q)$, $V_+=V(q')$. The short-time expansion assumes that we expand $W$ to power series in $\varepsilon$ to a given order, and calculate the appropriate coefficients using Eq.~(\ref{eq:eqW}). We could further expect that this results in coefficients depending on the potential $V(x)$ and its higher derivatives. However, this scheme is not complete, since the effective potential depends not only on the mid-point $x$, but also on the deviation $\bar x$, and the obtained equations for the coefficients cannot be solved in a closed form. In order to resolve this in a systematic way, we make use of the fact that, for short time of propagation, deviation $\bar x$ is on the average given by the diffusion relation $\bar x^2\propto\varepsilon$, allowing double expansion of $W$ in the form
\begin{equation}
\label{eq:ansatz}
W(x,\bar x;\varepsilon)=\sum_{m=0}^{\infty}\sum_{k=0}^{m}c_{m,k}(x)\,\varepsilon^{m-k}\bar x^{2k}\, .
\end{equation}
Restricting the above sum over $m$ to $p-1$ leads to level $p$ effective potential $W_p(x,\bar x;\varepsilon)$ which gives expansion of the effective action $S^*_p$ to order $\varepsilon^p$, and hence the level designation $p$ for both the effective action and the corresponding potential $W_p$. Thus, if the diffusion relation is applicable (which is always the case in Monte Carlo calculations), instead of the general double expansion in $\bar x$ and $\varepsilon$, we are able to obtain simpler, systematic expansion in $\varepsilon$ only. 

As shown previously \cite{prl-speedup, prb-speedup, pla-euler}, when used in Path Integral Monte Carlo simulations for calculation of long time amplitudes according to Eq.~(\ref{eq:AMC}), use of level $p$ effective action leads to the convergence of discretized amplitudes proportional to $\varepsilon^p$, i.e. as $1/N^p$, where $N$ is the number of time steps used in the discretization.

\begin{figure}[!t]
\centering
\includegraphics[width=6cm]{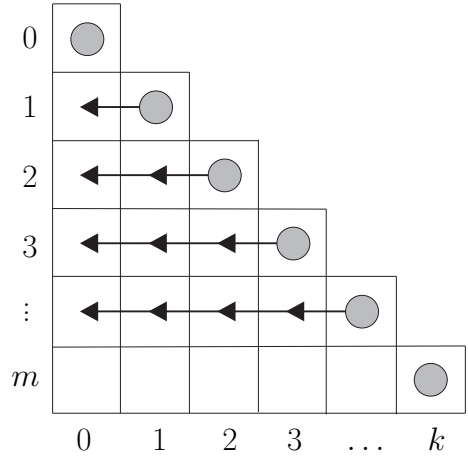}
\caption{Order in which the coefficients $c_{m,k}$ are calculated: diagonal ones
from Eq.~(\ref{eq:diagonal}), off-diagonal from recursion (\ref{eq:1Drec}).}
\label{fig:order}
\end{figure}

If we insert the above level $p$ expansion of the effective potential to Eq.(\ref{eq:eqW}), we obtain the recursion relation derived in Ref.~\cite{pre-recursive},
\begin{eqnarray}
8(m+k+1)\, c_{m,k}&=&(2k+2) (2 k+1)\, c_{m,k+1}
+c_{m-1,k}''-\sum_{l=0}^{m-2}\, \sum_{r} c_{l,r}'\,c_{m-l-2,k-r}'\nonumber\\
&&-\sum_{l=1}^{m-2}\,\sum_{r}2\,r(2k-2r+2)\,c_{l,r}\,c_{m-l-1,k-r+1}\, ,
\label{eq:1Drec}
\end{eqnarray}
where the sum over $r$ goes from ${\rm max}\{0,\ k-m+l+2\}$ to ${\rm min}\{k,\ l\}$. This recursion can be used to calculate all coefficients $c_{m,k}$ to a given level $p$, starting from the known initial condition, $c_{0, 0}=V$. The diagonal coefficients can be calculated immediately,
\begin{equation}
\label{eq:diagonal}
c_{m,m}=\frac{V^{(2m)}}{(2m+1)!}\, ,
\end{equation}
and for a given value of $m=0,\ldots p-1$, the coefficients $c_{m,k}$ follow recursively from evaluating (\ref{eq:1Drec}) for $k=m-1,\ldots,1,0$, as illustrated in Fig.~\ref{fig:order}.

\subsection{Extension to many-body systems}
\label{sec:mb}

The above outlined approach can be straightforwardly applied to many-body systems. Again the amplitude is expressed through the effective action and the corresponding effective potential, which now depends on mid-point positions and deviations of all particles. For simplicity, these vectors are usually combined into $D\times M$ dimensional vectors $\mathbf x$ and $\bar{\mathbf x}$, where $D$ is spatial dimensionality, and $M$ is the number of particles. In this notation,
\begin{equation}
A(\mathbf q, \mathbf q'; \varepsilon)=\frac{1}{(2\pi \varepsilon)^{DM/2}}\, e^{-\frac{2}{\varepsilon}\bar{\mathbf x}^2-\varepsilon W(\mathbf x, \bar{\mathbf x}; \varepsilon)}\, ,
\end{equation}
where initial and final position $\mathbf q=(\mathbf q_1, \ldots, \mathbf q_M)$ and $\mathbf q'=(\mathbf q'_1, \ldots, \mathbf q'_M)$ are analogously defined $D\times M$ dimensional vectors. Here we will not consider quantum statistics of particles. The required symmetrization or antisymmetrization must be applied after transition amplitudes are calculated using the effective potential.

Many-body transition amplitudes satisfy $D\times M$-dimensional generalization of the time-dependent Schr\" odinger equation, which leads to the equation for the effective potential similar to Eq.~(\ref{eq:eqW}), with vectors replacing previously scalar quantities,
\begin{equation}
W+\bar{\mathbf x}\cdot \bar{\boldsymbol\partial} W+\varepsilon\,\partial_\varepsilon W
-\frac{1}{8}\,\varepsilon\,\boldsymbol\partial^2 W-
\frac{1}{8}\,\varepsilon\,\bar{\boldsymbol\partial}^2 W
+\,\frac{1}{8}\,\varepsilon^2\,(\boldsymbol\partial W)^2
+\frac{1}{8}\,\varepsilon^2\,(\bar{\boldsymbol\partial} W)^2= \frac{1}{2}\, (V_++V_-)\, .
\label{eq:eqWMB}
\end{equation}
The effective potential for short-time amplitudes again can be written in the form of the double expansion in $\varepsilon$ and $\bar{\mathbf x}$. However, it turns out to be advantageous to use the expansion
\begin{equation}
\label{EXP}
W(x,\bar x;\varepsilon) = \sum_{m=0}^{\infty} \, \sum_{k=0}^{m}\varepsilon^{m-k}\,W_{m,k}(x,\bar x)\ ,
\end{equation}
and work with fully contracted quantities $W_{m,k}$
\begin{equation}
W_{m,k}(x,\bar x)= \bar x_{i_1} \bar x_{i_2} \cdots \bar x_{i_{2k}}c_{m,k}^{i_1,\ldots, i_{2k}}(x)\, ,
\end{equation}
rather than with the respective coefficients $c_{m,k}^{i_1,\ldots, i_{2k}}$. In this way we avoid the computationally expensive symmetrization over
all indices $i_1,\ldots, i_{2k}$. After inserting the above expansion into the equation for the effective potential, we obtain the recursion relation which represents a generalization of previously derived Eq.~(\ref{eq:1Drec}) for 1D case, and has the form
\begin{eqnarray}
\label{recursion}
8\, (m+k+1)\,W_{m,k}&=&\boldsymbol\partial^2 W_{m-1,k}+\bar{\boldsymbol\partial}^2 W_{m,k+1}
-\sum_{l=0}^{m-2}\,\sum_{r}(\boldsymbol\partial W_{l,r})\cdot(\boldsymbol\partial W_{m-l-2,k-r})\nonumber\\
&&-\sum_{l=1}^{m-2}\,\sum_{r}(\bar{\boldsymbol\partial} W_{l,r})\cdot(\bar{\boldsymbol\partial} W_{m-l-1,k-r+1})\, .
\label{eq:MBrec}
\end{eqnarray}
The sum over $r$ runs from ${\rm max}\{0,\ k-m+l+2\}$ to ${\rm min}\{k,\ l\}$, while diagonal quantities $W_{m,m}$ can be calculated directly,
\begin{equation}
\label{DIA}
W_{m,m}=\frac{1}{(2m+1)!}(\bar x\cdot{\boldsymbol\partial})^{2m}\,V\, .
\end{equation}
The above recursion disentangles, in complete analogy with the previously outlined case of one particle in 1D, and is solved in the order shown in Fig.~\ref{fig:order}.

\section{SPEEDUP Mathematica codes for deriving the higher-order effective actions}
\label{sec:Mathematica}

The effective action approach can be used for numerically exact calculation of short-time amplitudes if the effective potential $W_p$ can be analytically derived to sufficiently high values of $p$ such that the associated error is smaller than the required numerical precision. The error $\varepsilon^p$ for the effective action, obtained when level $p$ effective potential is used, translates into $\varepsilon^{p-DM/2}$ for a general many-body short-time amplitude. However, when amplitudes are calculated using the Path Integral Monte Carlo SPEEDUP C code \cite{speedup}, which will be presented in the next section, the errors of numerically calculated amplitudes are always proportional to $\varepsilon^p\sim 1/N^p$, where $N$ is number of time-steps in the discretization of the propagation time $T$.

Therefore, accessibility of higher-order effective actions is central to the application of this approach if it is used for direct calculation of short-time amplitudes \cite{ivanapre1, ivanapre2, becpla}, as well as in the case when PIMC code is used \cite{danicapla, jelapla, ivanapla}. However, increase in the level $p$ leads to the increase in complexity of analytic expressions for the effective potential. On one hand, this limits the maximal accessible level $p$ by the amount of memory required for symbolic derivation of the effective potential. On the other hand, practical use of large expressions for $W_p$ may slow down numerical calculations, and one can opt to use lower than the maximal available level $p$ when optimizing total CPU time required for numerical simulation. The suggested approach is to study time-complexity of the algorithm in practical applications, and to choose optimal level $p$ by minimizing the execution time required to achieve fixed numerical precision.

We have implemented efficient symbolic derivation of higher-order effective actions in Mathematica using the recursive approach. All source files described in this section are located in the {\tt Mathematica} directory of the SPEEDUP code distribution.

\subsection{General 1D Mathematica code}
\label{sec:M1D1Mathematica}

SPEEDUP code \cite{speedup} for symbolic derivation of the effective potential to specified level $p$ is implemented in Mathematica \cite{mathematica}, and is available in the {\tt EffectiveAction-1D.nb} notebook. It implements the algorithm depicted in Fig.~\ref{fig:order} and calculates the coefficients $c_{m,k}$ for $m=0,\ldots, p-1$ and $k=m,\ldots,0$, starting from the initial condition $c_{0,0}=V$. For a given value of $m$, the diagonal coefficient $c_{m,m}$ is first calculated from Eq.~(\ref{eq:diagonal}), and then all off-diagonal coefficients are calculated from the recursion (\ref{eq:1Drec}).

In this code the potential $V(x)$ is not specified, and the effective potential is derived for a general one-particle 1D theory. The resulting coefficients $c_{m,k}$ and the effective potential are expressed in terms of the potential $V$ and its higher derivatives. Level $p$ effective potential, constructed as
\begin{equation}
W_p(x,\bar x;\varepsilon)=\sum_{m=0}^{p-1}\sum_{k=0}^{m}c_{m,k}(x)\,\varepsilon^{m-k}\bar x^{2k}\, ,
\end{equation}
contains derivatives of $V$ to order $2p-2$.

The only input parameter of this Mathematica code is the level $p$ to which the effective potential should be calculated. As the code runs, it prints used amount of memory (in MB) and CPU time. This information can be used to estimate the required computing resources for higher values of $p$. The calculated coefficients can be exported to a file, and later imported for further numerical calculations. As an illustration, the file {\tt EffectiveAction-1D-export-p5.m} contains exported definition of all the coefficients $c_{m,k}$ calculated at level $p=5$, while the notebook {\tt EffectiveAction-1D-matching-p5.nb} contains matching output from the interactive session used to produce the above $p=5$ result.

The execution of this code on a typical 2 GHz CPU for level $p=10$ requires 10-15 MB of RAM and several seconds of CPU time. We have successfully run this code for levels as high as $p=35$ \cite{speedup}. SPEEDUP C code implements effective actions to the maximal level $p=18$, with the size of the corresponding C function around 2 MB. If needed, higher levels $p$ can be easily implemented in C and added to the existing SPEEDUP code.

\subsection{General 2D and 3D Mathematica code}
\label{sec:M1D2D3Mathematica}

Although we have developed Mathematica code capable of deriving effective actions for a general many-body theory in arbitrary number of spatial dimensions, in practical applications in 2D and 3D it can be very advantageous to use simpler codes, able to produce results to higher levels $p$ than the general code \cite{becpla, ivanapre2}.

This is done in files {\tt EffectiveAction-2D.nb} and {\tt EffectiveAction-3D.nb}, where the recursive approach is implemented directly in 2D and 3D. Execution of these codes requires more memory: for $p=10$ effective action one needs 60 MB in 2D case, while in 3D case the needed amount of memory increases to 860 MB. On the other hand, the execution time is several minutes for 2D code and around 30 minutes for 3D code.

The distribution of the SPEEDUP code contains exported $p=5$ definitions of contractions $W_{m, k}$ for both 2D and 3D general potential, as well as matching outputs from interactive sessions used to generate these results.

\subsection{Model-specific Mathematica codes}
\label{sec:D1modelsMathematica}

When general expressions for the effective actions, obtained using the above described SPEEDUP Mathematica codes, are used in numerical simulations, one has to specify the potential $V$ and its higher derivatives to order $2p-2$ in order to be able to calculate transition amplitudes. Such approach is justified for systems where the complexity of higher derivatives increases. However, for systems where this is not the case, or where only a limited number of derivatives is non-trivial (e.g. polynomial interactions), it might be substantially beneficial to specify the potential at the beginning of the Mathematica code and calculate the derivatives explicitly when iterating the recursion.

Using this approach, one is able to obtain coefficients $c_{m,k}$ and the effective potential $W$ directly as functions of the mid-point $x$. This is implemented in the notebooks {\tt EffectiveAction-1D-AHO.nb} and {\tt EffectiveAction-2D-AHO.nb} for the case of anharmonic oscillators in 1D and 2D,
\begin{eqnarray}
&&\hspace*{-1.2cm}
V_{1D-AHO}(x)=\frac{A}{2}\, x^2+\frac{g}{24}\,x^4\, ,\\
&&\hspace*{-1.2cm}
V_{2D-AHO}(x)=\frac{A}{2}\, (x^2+y^2)+\frac{g}{24}\, (x^2+y^2)^2\, .
\end{eqnarray}
These codes can be easily executed within few seconds and with the minimal amounts of memory even for $p=20$. For 1D anharmonic oscillator we have successfully calculated effective actions to excessively large value $p=144$, and in 2D to $p=67$ \cite{speedup}, to illustrate the advantage of this model-specific method.

Similar approach can be also used in another extreme case, when the complexity of higher derivatives of the potential $V$ increases very fast, so that entering the corresponding expressions to the code becomes impractical. Even in this situation expressions for effective actions can be usually simplified using some appropriate model-specific ansatz. The form of such ansatz can be deduced from the form of model-specific effective potentials, and then used to simplify their derivation. Such use-case is illustrated in the SPEEDUP Mathematica code for the modified P\" oscl-Teller potential,
\begin{equation}
V_{1D-MPT}(x)=-\frac{\lambda}{(\cosh \alpha x)^2}\, .
\label{eq:1D-MPT}
\end{equation}
For this potential, the coefficients $c_{m,k}$ of the effective potential can be expressed in the form
\begin{equation}
c_{m, k}(x)=\sum_{l=0}^m d_{m, k, l}\, \frac{(\tanh \alpha x)^{2l}}{(\cosh \alpha x)^{2m-2l+2}}\, ,
\label{eq:1D-MPTansatz}
\end{equation}
and newly introduced constant coefficients $d_{m,k,l}$ can be calculated using the model-speci\-fic recursion in {\tt EffectiveAction-1D-MPT.nb}. The form of the ansatz (\ref{eq:1D-MPTansatz}) is deduced from the results of executing general 1D Mathematica code, with the model-specific potential (\ref{eq:1D-MPT}) defined before the recursion calculation of the coefficients is performed. Using this approach, we were able to obtain maximal level $p=41$ effective action \cite{speedup}.

\subsection{General many-body Mathematica code}
\label{sec:mbMathematica}

SPEEDUP Mathematica code for calculation of effective action for a general many-body theory is implemented using the MathTensor \cite{mathtensor} package for tensorial calculations in Mathematica. This general implementation required some new functions related to the tensor calculus to be defined in the source notebook {\tt EffectiveAction-ManyBody.nb} provided with the SPEEDUP code.

The function {\tt GenNewInd[n]} generates the required number {\tt n} of upper and lower indices using the MathTensor function {\tt UpLo}, with the assigned names {\tt up1}, {\tt lo1}, \ldots, as well as lists {\tt upi} and {\tt loi}, each containing {\tt n} strings corresponding to the names of generated indices. These new indices are used in the implementation of the recursion for calculation of derivatives of $W_{m,k}$, contractions of the effective potential, and for this reason had to be explicitly named and properly introduced.

The expressions obtained by iterating the recursion contain large numbers of contractions, and function {\tt NewDefUnique[contr]} replaces all contracted indices with the newly-introduced dummy ones in the contraction {\tt contr}, so that they do not interfere with the calculation of derivatives in the recursion. This is necessary since the derivatives in recursion do not distinguish contracted indices from non-contracted ones if their names happen to be generated by the function {\tt GenNewInd}. Note that the expression {\tt contr} does not have to be full contraction, i.e. function {\tt NewDefUnique} will successfully act on tensors of any kind if they have contracted indices, while it will leave them unchanged if no contractions are present.

The function {\tt NewDerivativeVec[contr, vec, ind]} implements calculation of the first derivative of the tensor {\tt contr} (which may or may not contain contracted indices, but if it does, they are supposed to be uniquely defined dummy ones, which is achieved using the function {\tt NewDefUnique}). The derivative is calculated with respect to vector {\tt vec} with the vectorial index {\tt ind}. The index {\tt ind} can be either lower or upper one, and has to be generated previously by the function {\tt GenNewInd}.

Finally, the function {\tt NewLaplacianVec[contr, vec]} implements the Laplacian of the tensor {\tt contr} with respect to the vector {\tt vec}, i.e. it performs the calculation of contractions of the type
\begin{equation}
\frac{\partial}{\partial\mathtt{vec}_i}\,\frac{\partial}{\partial\mathtt{vec}^i}\,\, \mathtt{contr}\, .
\end{equation}

After all described functions are defined, the execution of the code proceeds by setting the desired level of the effective action {\tt p}, generating the needed number of named indices using the function call {\tt GenNewInd[2 p + 2]}, and then by performing the recursion according to the scheme illustrated in Fig.~\ref{fig:order}. The use of MathTensor function {\tt CanAll} in the recursion ensures that the obtained expressions for {\tt W[m, k]} will be simplified if possible. This is achieved in MathTensor by sorting and renaming all dummy indices using the same algorithm and trying to simplify the expression obtained in such way. By default, Mathematica will distinguish contracted indices in two expressions if they are named differently, and MathTensor works around it using the renaming scheme implemented in {\tt CanAll}.

The computing resources required for the execution of the many-body SPEEDUP Mathematica code depend strongly on the level of the effective action. For example, for level $p=5$ the code can be run within few seconds with the minimal memory requirements. The notebook with the matching output of this calculation is available as {\tt EffectiveAction-ManyBody-matching-p5.nb}, and the exported results for {\tt W[m, k]} are available in {\tt EffectiveAction-ManyBody-export-p5.m}. We were able to achieve maximal level $p=10$ \cite{speedup}, with the CPU time of around 2 days on a recent 2 GHz processor. The memory used by Mathematica was approximately 1.6 GB.

Note that exporting the definition of the effective potential from Mathematica to a file will yield lower and upper indices named {\tt ll1}, {\tt uu1}, etc. In order to import previous results and use them for further calculations with the provided Mathematica code, it is necessary to replace indices in the exported file to the proper index names used by the function {\tt GenNewInd}. This is easily done using find/replace feature of any text editor. Prior to importing definition of the effective potential, it is necessary to initialize MathTensor and all additional functions defined in the notebook {\tt EffectiveAction-ManyBody.nb}, and to generate the needed number of named indices using the function call {\tt GenNewInd[2p+2]}.

\section{SPEEDUP C codes for Monte Carlo calculation of 1D transition amplitudes}
\label{sec:C}

For short times of propagation, the effective actions derived using the above described Mathematica codes can be directly used. This has been extensively used in Refs.~\cite{ivanapre1, ivanapre2}, where SPEEDUP codes were applied for numerical studies of several lower-dimensional models and calculation of large number of energy eigenvalues and eigenfunctions. The similar approach is used in Ref.~\cite{becpla}, where SPEEDUP code was used to study properties of fast-rotating Bose-Einstein condensates in anharmonic trapping potentials. The availability of a large number of eigenstates allowed not only precise calculation of global properties of the condensate (such as condensation temperature and ground state occupancy), but also study of density profiles and construction of time-of-flight absorption graphs, with the exact quantum treatment of all available eigenfunctions.

However, in majority of applications the time of propagation cannot be assumed to be small. The effective actions are found to have finite radius of convergence \cite{ivanapre1}, and if the typical propagation times in the considered case exceed this critical value, Path Integral Monte Carlo approach must be used in order to accurately calculate the transition amplitudes and the corresponding expectation values \cite{danicapla, jelapla}. As outlined earlier, in this case the time of propagation $T$ is divided into $N$ time steps, such that $\varepsilon=T/N$ is sufficiently small and that the effective action approach can be used. The discretization of the propagation time leads to the following expression for the discretized amplitude
\begin{equation}
A_N^{(p)}(a,b;T)=\int\frac{dq_1\cdots dq_{N-1}}{(2\pi\varepsilon)^{N/2}}\, e^{-S_N^{(p)}}\, ,
\end{equation}
where $S_N^{(p)}$ stands for the discretized level $p$ effective action,
\begin{equation}
S_N^{(p)}=\sum_{k=0}^{N-1}\left[ \frac{(q_{k+1}-q_k)^2}{2\varepsilon}+\varepsilon\, W_p(x_k, \bar x_k; \varepsilon)\right],
\label{eq:SNp}
\end{equation}
and $q_0=a$, $q_N=b$, $x_k=(q_{k+1}+q_k)/2$, $\bar x_k=(q_{k+1}-q_k)/2$.

Level $p$ discretized effective action is constructed from the corresponding effective potential $W_p$, calculated as power series expansion to order $\varepsilon^{p-1}$. Since it enters the action multiplied by $\varepsilon$, this leads to discretized actions correct to order $\varepsilon^p$, i.e. with the errors of the order $\varepsilon^{p+1}$. The long-time transition amplitude $A_N^{(p)}(a,b;T)$ is a product of $N$ short-time amplitudes, and its errors are expected to scale as $N\cdot\varepsilon^{p+1}\sim 1/N^p$, as has been shown in Refs.~\cite{prl-speedup, prb-speedup, pla-euler, ivanapla} for transition amplitudes, and in Refs.~\cite{jelapla, ivanapla} for expectation values, calculated using  the corresponding consistently improved estimators.

\subsection{Algorithm and structure of the code}
\label{sec:algorithm}

SPEEDUP C source is located in the {\tt src} directory of the code distribution  \cite{speedup}. It uses the standard Path Integral Monte Carlo algorithm for calculation of transition amplitudes. The trajectories are generates by the bisection algorithm \cite{ceperley}, hence the number of time-steps $N$ is always given as a power of two, $N=2^s$. When the amplitude is calculated with $2^s$ time steps, we can also easily calculate all discretized amplitudes in the hierarchy $2^{s-1}$, \ldots, $2^0$ at no extra cost. This requires only minor additional CPU time and memory, since the needed trajectories are already generated as subsets of maximal trajectories with $2^s$ time-steps.

The trajectory is constructed starting from bisection level $n=0$, where we only have initial and final position of the particle. At bisection level $n=1$ the propagation is divided into two time-steps, and we have to generate coordinate $q$ of the particle at the moment $T/2$, thus constructing the piecewise trajectory connecting points $a$ at the time $t=0$, $q$ at $t=T/2$, and $b$ at $t=T$. The coordinate $q$ is generated from the Gaussian probability density function centered at $(a+b)/2$ and with the width $\sigma_1=\sqrt{T/2}$. The procedure continues iteratively, and each time a set of points is added to the piecewise trajectory. At each bisection level $n$ the coordinates are generated from the Gaussian centered at mid-point of coordinates generated at level $n-1$, with the width $\sigma_n=\sqrt{T/2^n}$. To generate numbers $\eta$ from the Gaussian centered at zero we use Box-M\" uller method,
\begin{equation}
\eta = \sqrt{-2\sigma_n^2\ln\xi_1}\, \cos 2\pi\xi_2\, ,
\end{equation}
where numbers $\xi_1$ and $\xi_2$ are generated from the uniform distribution on the interval $[0, 1]$, using the SPRNG library \cite{sprng}. If the target bisection level is $s$, then at bisection level $n\leq s$ we generate $2^{n-1}$ numbers using the above formula, and construct the new trajectory by adding to already existing points the new ones, according to
\begin{equation}
q[(1+2i)\cdot 2^{s-n}]=\eta_i+\frac{q[i\cdot 2^{s-n+1}]+q[(i+1)\cdot 2^{s-n+1}]}{2}\, ,
\end{equation}
where $i$ runs from 0 to $2^{n-1}-1$. This ensures that at bisection level $s$ we get trajectory with $N=2^s$ time-steps, consisting of $N+1$ points, with boundary conditions $q[0]=a$ and $q[N]=b$. At each lower bisection level $n$, the trajectory consists of $2^n+1$ points obtained from the maximal one (level $s$ trajectory) as a subset of points $q[i\cdot 2^{s-n}]$ for $i=0,1,\ldots, 2^n$.

The use of trajectories generated by the bisection algorithm requires normalization factors from all Gaussian probability density functions with different widths to be taken into account. This normalization is different for each bisection level, but can be calculated easily during the initialization phase.

The basic C code is organized in three source files, {\tt main.c}, {\tt p.c} and {\tt potential.c}, with the accompanying header files. The file {\tt potential.c} (its name can be changed, and specified at compile time) must contain a user-supplied function {\tt V0()}, defining the potential $V$. For a given input value of the coordinate, {\tt V0()} should initialize appropriate variables to the value of the potential $V$ and its higher derivatives to the required order $2p-2$. When this file is prepared, SPEEDUP code can be compiled and used. The distributed source contains definition of 1D-AHO potential in the file {\tt potential.c}, the same as in the file {\tt 1D-AHO.c}.

The execution of the SPEEDUP code starts with the initialization and allocation of memory in the {\tt main()} function, and then the array of amplitudes and associated MC error estimates for each bisection level $n=0,\ldots, s$ is calculated by calling the function {\tt mc()}. After printing the output, {\tt main()} deallocates used memory and exits. Function {\tt mc()} which implements the described MC algorithm is also located in the file {\tt main.c}, as well as the function {\tt distr()}, which generates maximal (level $s$) trajectories.

The function {\tt mc()} contains main MC sampling loop. In each step new level $s$ trajectory is generated by calling the function {\tt distr()}. Afterwards, for each bisection level $n$, function {\tt func()} is invoked. This function is located in the file {\tt p.c}, and returns the value of the function $e^{-S}$, properly normalized, as described earlier. This value (and its square) is accumulated in the MC loop for each bisection level $n$ and later averaged to obtain the estimate of the corresponding discretized amplitude and the associated MC error.

The function {\tt func()} makes use of C implementation of earlier derived effective actions for a general 1D potential. For a given trajectory at the bisection level $n$, {\tt func()} will first initialize appropriate variables with the values of the potential and its higher derivatives (to the required level $2p-2$) by calling the user-supplied function {\tt V0()}, located in the file {\tt potential.c}. Afterwards the effective action is calculated according to Eq.~(\ref{eq:SNp}), where the effective potential is calculated by the function {\tt Wp()}, located in the file {\tt p.c}. The desired level $p$ of the effective action is selected by defining the appropriate pre-processor variable when the code is compiled.

In addition to this basic mode, when SPEEDUP code uses general expression for level $p$ effective action, we have also implemented model-specific mode, described earlier. If effective actions are derived for a specific model, then user can specify an alternative {\tt p.c} file to be used within the directory {\tt src/models/<model>}, where {\tt <model>} corresponds to the name of the model. If this mode is selected at compile time, the compiler will ignore {\tt p.c} from the top {\tt src} directory, and use the model-specific one, defined by the user. The distributed source contains model definitions for 1D-AHO and 1D-MPT potentials in directories {\tt src/models/1D-AHO} and {\tt src/models/1D-MPT}. Note that in this mode the potential is specified directly in the definition of the effective potential, and therefore the function {\tt V0()} is not used (nor the {\tt potential.c} file).

\subsection{Compiling and using SPEEDUP C code}
\label{sec:compiling}

SPEEDUP C source can be easily compiled using the {\tt Makefile} provided in the top directory of the distribution. The compilation has been thoroughly tested with GNU, Intel and IBM XLC compilers. In order to compile the code one has to specify the compiler which will be used in the {\tt Makefile} by setting appropriately the variable {\tt COMPILER}, and then to proceed with the standard command of the type {\tt make <target>}, where {\tt <target>} could be one of {\tt all}, {\tt speedup}, {\tt sprng}, {\tt clean-all}, {\tt clean-speedup}, {\tt clean-sprng}.

The SPRNG library \cite{sprng} is an external dependency, and for this reason it is located in the directory {\tt src/deps/sprng4.0}. In principle, it has to be compiled only once, after the compiler has been set. This is achieved by executing the command {\tt make sprng}. Afterwards the SPEEDUP code can be compiled and easily linked with the already compiled SPRNG library. Note that if the compiler is changed, SPRNG library has to be recompiled with the same complier in order to be successfully linked with the SPEEDUP code.

To compile the code with level $p=10$ effective action and user-supplied function {\tt V0()} located in the file {\tt src/1D-AHO.c}, the following command can be used:\\
\hspace*{6mm}{\tt make speedup P=10 POTENTIAL=1D-AHO.c}\\
If not specified, {\tt POTENTIAL=potential.c} is used, while the default level of the effective action is {\tt P=1}. To compile the code using a model-specific definition of the effective potential, instead of the {\tt POTENTIAL} variable, we have to appropriately set the {\tt MODEL} variable on the command line. For example, to compile the supplied {\tt p.c} file for 1D-MPT model located in the directory {\tt src/models/1D-MPT} using the level $p=5$ effective action, the following command can be used:\\
\hspace*{11mm}{\tt make speedup P=5 MODEL=1D-MPT}\\
All binaries compiled using the {\tt POTENTIAL} mode are stored in the {\tt bin} directory, while the binaries for the {\tt MODEL} mode are stored in the appropriate {\tt bin/models/<model>} directory. This information is provided by the {\tt make} command after each successful compilation is done.

The compilation is documented in more details in the supplied {\tt README.txt} files. The distribution of the SPEEDUP code also contains examples of compilation with the GNU, Intel and IBM XLC compilers, as well as matching outputs and results of the execution for each tested compiler, each model, and for a range of levels of the effective action $p$.

\begin{figure}[!b]
\centering
\includegraphics[width=9.5cm]{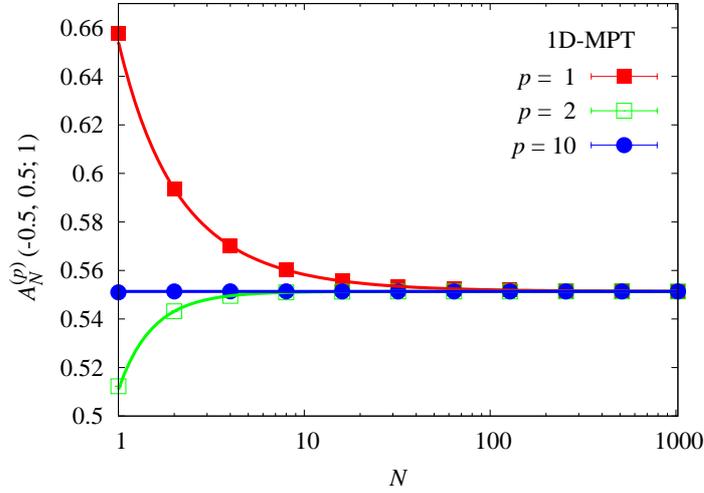}
\caption{Convergence of SPEEDUP Monte-Carlo  results for the transition amplitude $A_N^{(p)}(-0.5, 0.5; 1)$ of 1D-MPT potential as a function of the number of time steps $N$, calculated with level $p=1, 2, 10$ effective actions, with the parameters of the potential $\lambda=\alpha=1$. The full lines give the fitted functions (\ref{eq:fit}), where the constant term  $A_p$ corresponds to the continuum-theory amplitude $A(-0.5, 0.5; 1)$. The number of Monte-Carlo samples was $N_{\rm MC}=10^{6}$.}
\label{fig:conv}
\end{figure}

Once compiled, the SPEEDUP code can be used to calculate long-time amplitudes of a system in the specified potential $V$. If executed without any command-line arguments, the binary will print help message, with details of the usage. The obligatory arguments are time of propagation {\tt T}, initial and final position {\tt a} and {\tt b}, maximal bisection level {\tt s}, number of MC samples {\tt Nmc} and {\tt seed} for initialization of the SPRNG random number generator. All further arguments are converted to numbers of the {\tt double} type and made available in the array {\tt par} to the function {\tt V0()}, or to the model-specific functions in the file {\tt src/models/<model>/p.c}. The output of the execution contains calculated value of the amplitude for each bisection level $n=0,\ldots,s$ and the corresponding MC estimate of its error (standard deviation). At bisection level $n=0$, where no integrals are actually calculated and the discretized $N=1$ amplitude is simply given by an analytic expression, zero is printed as the error estimate.

\begin{figure}[!b]
\centering
\includegraphics[width=7.2cm]{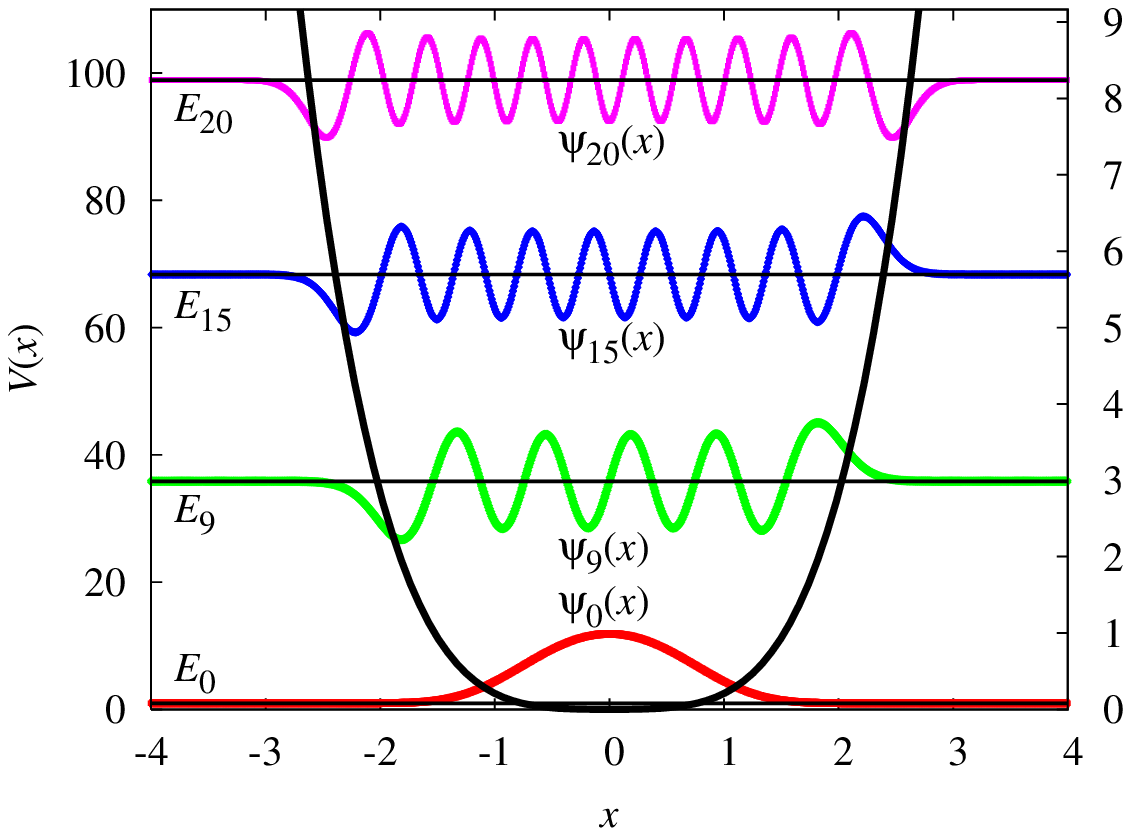}
\includegraphics[width=7.2cm]{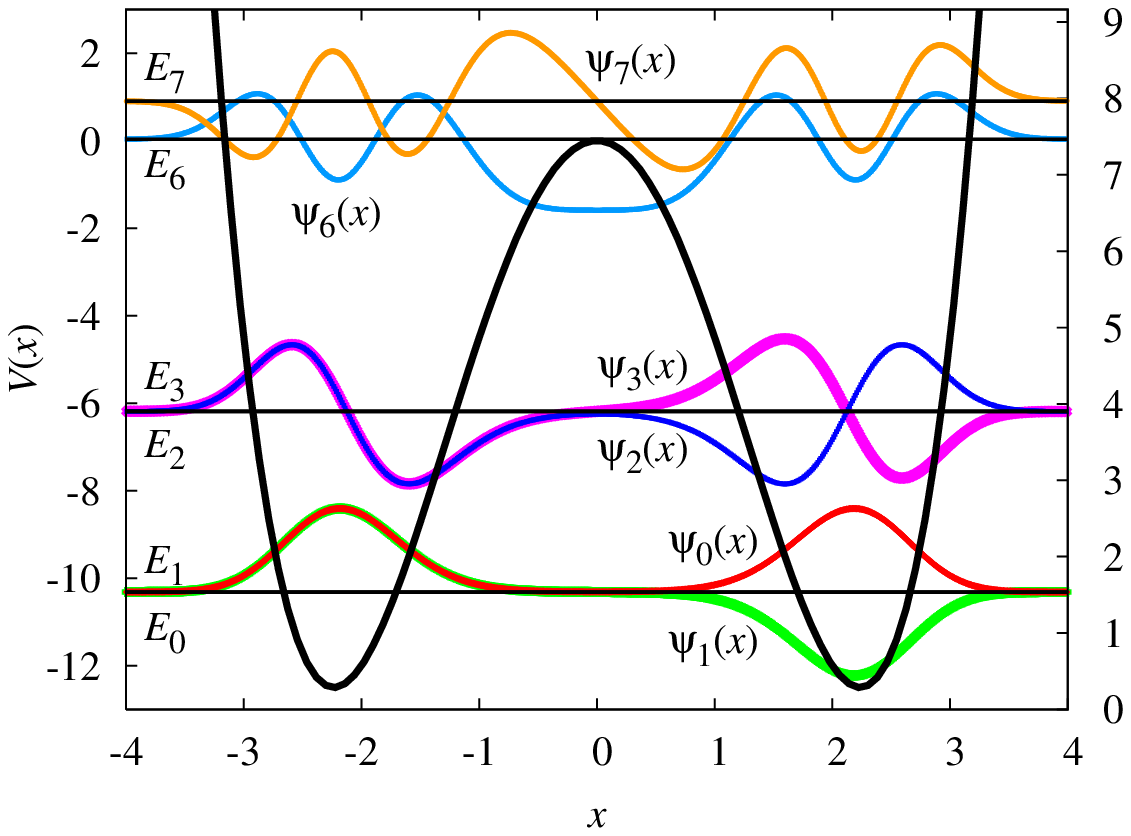}
\caption{(left) The anharmonic potential 1D-AHO, its energy eigenvalues (horizontal lines) and eigenfunctions, obtained by direct diagonalization of the space-discretized matrix of the evolution operator with level $p=21$ effective action and parameters $A=1$, $g=48$. The discretization cutoff was $L=8$, spacing $\Delta=9.76\cdot 10^{-4}$, and time of propagation $t=0.02$. (right) Results for the double-well potential, $A=-10$, $g=12$, $L=10$, $\Delta=1.22\cdot 10^{-3}$, $t=0.1$. On both graphs, left $y$-axis corresponds to $V(x)$ and energy eigenvalues, while scale on the right $y$-axis corresponds to values of eigenfunctions, each vertically shifted to level with the appropriate eigenvalue.}
\label{fig:phi4states}
\end{figure}

Fig.~\ref{fig:conv} illustrates the typical results obtained from the SPEEDUP code on the example of 1D-MPT theory. In this figure we can see the convergence of numerically calculated amplitudes with the number of time-steps $N$ to the exact continuum value, obtained in the limit $N\to\infty$. Such convergence is obtained for each level $p$ of the effective action used. However, the convergence is much faster when higher-order effective action is used. Note that all results corresponding to the one value of level $p$ on the graph are obtained from a single run of the SPEEDUP code with the maximal bisection level $s=10$. The simplest way to estimate the continuum value of the amplitude is to fit numerical results from single run of the code to the appropriate level $p$ fitting function \cite{prl-speedup, prb-speedup, pla-euler},
\begin{equation}
A_N^{(p)}=A^{(p)}+\frac{B^{(p)}}{N^p}+\frac{C^{(p+1)}}{N^{p+1}}+\cdots
\label{eq:fit}
\end{equation}
The constant term obtained by fitting corresponds to the best estimate of the exact amplitude which can be found from the available numerical results.

As mentioned earlier, the effective action approach can be used for accurate calculation of a large number of energy eigenstates and eigenvalues by diagonalization of the space-discretized matrix of transition amplitudes \cite{sethia1990, sethia1999, ivanapre1, ivanapre2, becpla}. Fig.~\ref{fig:phi4states} illustrates this for the case of an anharmonic and double-well potential. The graph on the left gives several eigenvalues and eigenstates for 1D-AHO potential with $A=1$ and quartic anharmonicity $g=48$, while the graph on the right gives low-lying spectrum and eigenfunctions of the double-well potential, obtained for $A=-10$, with the moderate anharmonicity $g=12$. More details on this approach, including study of all errors associated with the discretization process, can be found in Refs.~\cite{ivanapre1, ivanapre2}.

\section{Conclusions}
\label{sec:conclusions}

In this paper we have presented SPEEDUP Mathematica and C codes, which implement the effective action approach for calculation of quantum mechanical transition amplitudes. The developed Mathematica codes provide an efficient tool for symbolic derivation of effective actions to high orders for specific models, for a general 1D, 2D and 3D single-particle theory, as well as for a general many-body systems in arbitrary number of spatial dimensions. The recursive implementation of the code allows symbolic calculation of extremely high levels of effective actions, required for high-precision calculation of transition amplitudes.

For calculation of long-time amplitudes we have developed SPEEDUP C Path Integral Monte Carlo code. The C implementation of a general 1D effective action to maximal level $p=18$ and model-specific effective actions provide fast $1/N^p$ convergence to the exact continuum amplitudes.

Further development of the SPEEDUP C codes will include parallelization using MPI, OPENMP and hybrid programming model, C implementation of the effective potential to higher levels $p$, as well as providing model-specific effective actions for relevant potentials, including many-body systems.

\section*{Acknowledgements}
The authors gratefully acknowledge useful discussions with Axel Pelster and Vladimir Slavni\' c.
This work was supported in part by the Ministry of Education and Science of the Republic of Serbia, under project No. ON171017, and bilateral project NAD-BEC funded jointly with the German Academic Exchange Service (DAAD), and by the European Commission under EU FP7 projects PRACE-1IP, HP-SEE and EGI-InSPIRE.

\end{document}